\documentclass[]{aa}     % LaTeX A&A  Standard Fonts
\usepackage{graphics}
\def\cc{\,{\rm cm^{-3}}}
\def\cm2{\,{\rm cm^{-2}}}
\def\kms{\,{\rm {km\,s^{-1}}}}
\def\kkms{\,{\rm {K\,km s^{-1}}}}
\def\co{\,{\rm ^{12}CO}}
\def\13co{\,{\rm ^{13}CO}}
\def\h2{\,{\rm H_{2}}}
\def\Msun{\rm M_{\odot}}

\def\aua{{\rm A\&A} }
\def\auas{{\rm A\&AS} }
\def\apj{{\rm ApJ} }
\def\aj{{\rm AJ} }

\def\apjl{{\rm ApJL} }

\def\mnras{{\rm MNRAS} }
\def\pasj{{\rm PASJ} }
\def\pasp{{\rm PASP} }
\begin{document}
 
\title{Molecular gas in compact galaxies}
 
   \subtitle{}
 
\author{F.P. Israel
          \inst{1}
           }
 
   \offprints{F.P. Israel}
 
  \institute{Sterrewacht Leiden, P.O. Box 9513, 2300 RA Leiden,
             The Netherlands}
 
\date{
Received ????; accepted ????}
 
\abstract{New observations of eleven compact galaxies in the $\co$ $J
= 2-1$ and $J = 3-2$ transitions are presented.  From these
observations and literature data accurate line ratios in matched beams
have been constructed, allowing the modelling of physical parameters.
Matching a single gas component to observed line ratios tends to
produce physically unrealistic results, and is often not possible at
all.  Much better results are obtained by modelling two distinct gas
components.  In most observed galaxies, the molecular gas is warm
($T_{\rm k} = 50 - 150$ K) and at least partially dense ($n(\h2) \geq
3000 \cc$). Most of the gas-phase carbon in these galaxies is in
atomic form; only a small fraction ($\sim 5\%$) is in carbon
monoxide. Beam-averaged CO column densities are low (of the order of
$10^{16}\,\cm2$). However, molecular hydrogen column densities are
high (of the order of $10^{22} \,\cm2$) confirming large CO-to-$\h2$
conversion factors (typically $X = 10^{21} - 10^{22} \cm2/\kkms$)
found for low-metallicity environments by other methods. From CO
spectroscopy, three different types of molecular environment may be
distinguished in compact galaxies. Type I (high rotational and
isotopic ratios) corresponds to hot and dense molecular clouds
dominated by star-forming regions. Type II has lower ratios, similar
to the mean found for infrared-luminous galaxies in general, and
corresponds to environments engaged in, but not dominated by,
star-forming activity. Type III, characterized by low $\co$
(2--1)/(1--0) ratios, corresponds to mostly inactive environments of
relatively low density.   }

\maketitle
 
\section{Introduction}

Compared to large spiral galaxies, dwarf and compact galaxies are
difficult to detect in molecular lines, even in those of the
relatively abundant CO molecule. However, molecular clouds constitute
the unique environment out of which the stars are formed and knowledge
of their occurrence, spatial distribution, mass and physical condition
is essential to gain insight in the process of star formation in
galaxies. For this reason, many surveys of dwarf and compact galaxies
have been conducted over the last decades, although generally with low
detection rates (Table\,\ref{survey}).

%Table 1
\begin{table}
\caption[]{CO surveys of dwarf and compact galaxies}
\begin{center}
\begin{tabular}{lrrr}
\hline
\noalign{\smallskip}
Reference                    &\multicolumn{2}{c}{Number}&Line\\
                             &Obs. &Det. & \\
\noalign{\smallskip}
\hline
\noalign{\smallskip}
Elmegreen et al. (1980)      &  7& 1& 1-0\\
Gordon et al. (1982)         &  7& 0& 1-0\\
Israel $\&$ Burton (1986)    & 12& 0& 1-0\\
Thronson $\&$ Bally (1987)   & 22&10& 1-0\\
Tacconi $\&$ Young (1987)    & 15& 6& 1-0\\
Arnault et al. (1988)        & 12& 0& 1-0\\
Sage et al. (1992)           & 15& 8& 1-0/2-1\\
Hunter $\&$ Sage (1993)      &  5& 0& 1-0/2-1\\
Israel et al. (1995)         & 25& 6& 1-0\\
Gondhalekar et al. (1998)    & 29& 5& 1-0\\
Taylor et al. (1998)         & 11& 3& 1-0\\
Barone et al. (2000)         & 10& 2& 1-0/2-1 \\
Meier et al. (2001)          &  8& 8& 3-2\\
Braine et al. (2001)         & 10& 8& 1-0/2-1 \\
Boselli et al. (2002)        &  6& 1& 1-0\\
Albrecht et al. (2004)       & 64&41& 1-0/2-1 \\
Leroy et al. (2005)          &121&47& 1-0\\
\noalign{\smallskip}
\hline
\end{tabular}
\end{center}
\label{survey}
\end{table}
  
Detection rates are highest for `tidal' dwarfs (cf. Braine et al.
2001), `big' dwarfs (Sm) and galaxies that are only modestly
metal-poor.  They are lowest for `small' dwarfs (Im), blue compact
dwarf galaxies (BCDGs) and practically non-existent for very
metal-poor dwarfs (Taylor et al. 1998; Barone et al. 2000).  In the
extensive survey by Albrecht et al. (2004), detection rates are $72\%$
and $39\%$ for Sm and Im galaxies respectively (see also Leroy et al.,
2005).

Velocity-integrated $\co(1-0)$ fluxes are often combined with the
so-called CO-to-H$_{2}$ conversion factor $X$ to derive a beam-averaged
$\h2$ column-density, hence an estimate for the $\h2$ mass in the
beam. It is widely (but not yet universally) accepted that this factor
$X$ is a function of environmental conditions (such as metallicity,
radiation density etc.). However, the extent to which $X$ is sensitive
to e.g. changes in metallicity is still a matter of debate. Arguments
for a strong dependence of $X$ on metallicity as traced by the oxygen
abundance [O]/[H] have been presented by Israel (1997; 2000), Barone
et al. (2000) and Boselli et al. (2002).

Obviously, it is much preferable to determine actual molecular column
densities and masses from an analysis of the physical condition of the
gas than from an empirically determined factor that relates column
density to the flux of an optically thick line.  In principle,
multi-line observations can be used as a diagnostic tool to estimate
the physical condition of the gas, but this requires that the various
transitions have been observed at the same celestial position with
matched beamsizes.  A glance at Table\,\ref{survey} shows that only a
few surveys sample CO line transitions other than $J$=1--0. In a
number of surveys, $J$=2--1 and $J$=1--0 measurements were made
simultaneously with the same telescope, therefore with beam areas
differing by a factor of four. Mapping the $J$=2--1 line over the
extent of the $J$=1--0 beam would make it possible to compare the two
transitions, but this is rarely done. Only the very recent survey by
Albrecht et al. combines measurements in the two transitions obtained
with the IRAM 30m and the SEST 15m apertures, providing matched (SEST)
$J$=2--1 and (IRAM) $J$=1--0 measurements. In all preceding surveys,
including the CSO $\co(3-2)$ survey by Meier et al. 2001, line ratios
could only be estimated as a function of (assumed) source extent
(point source, extended source, or inbetween). As can easily be seen,
this introduces very large uncertainties, effectively ruling out
meaningful modelling of physical paramters.

In this paper we present new observations, which together with 
measurements published in the literature, allow us for the first
time to construct reliable line ratios in matched beams for
a sample of about a dozen dwarf or compact galaxies.

%Figure1: Central Profiles
\begin{figure*}[]
\resizebox{17cm}{!}{\rotatebox{270}{\includegraphics*{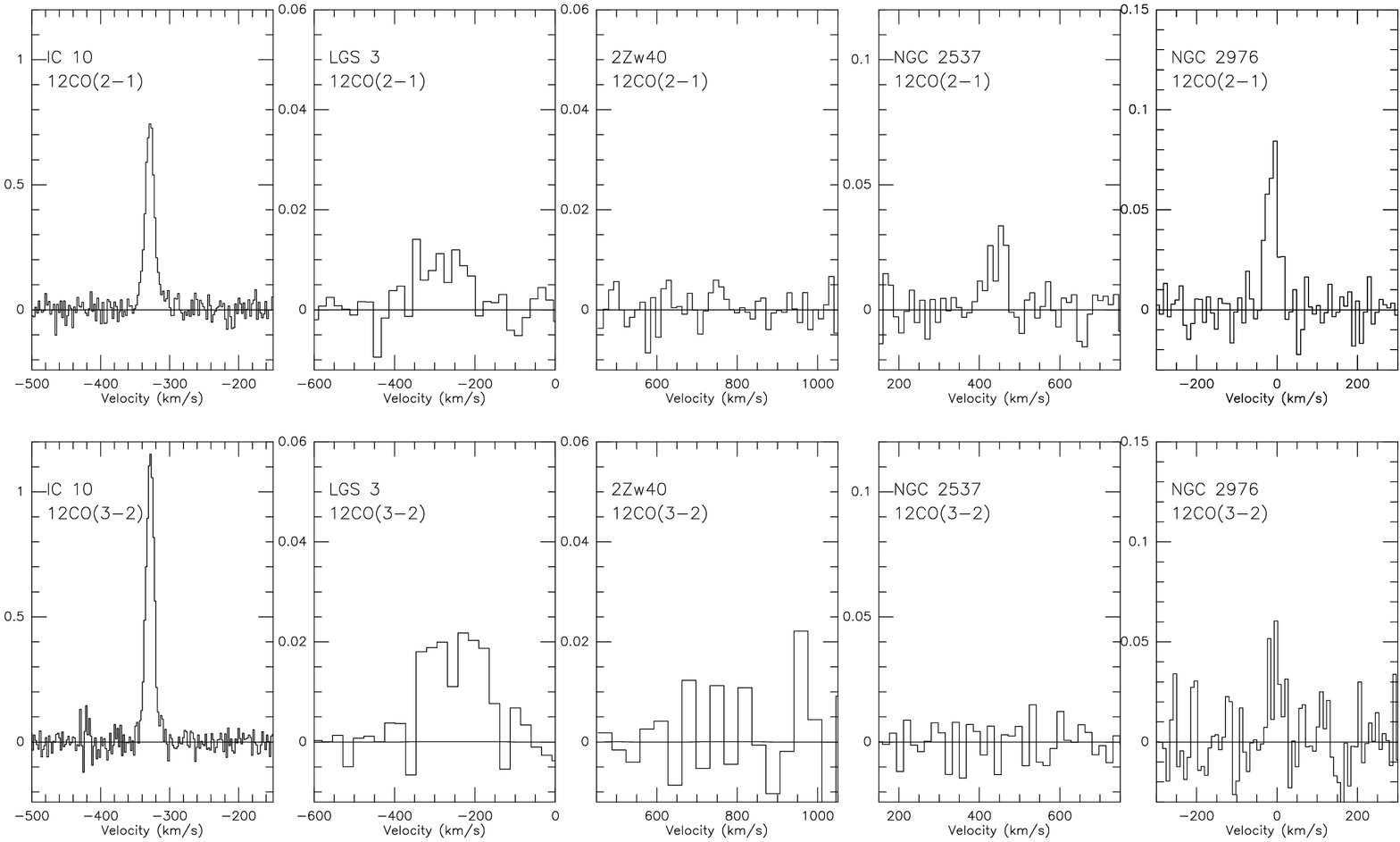}}}

\vspace{0.3cm}
\resizebox{17cm}{!}{\rotatebox{270}{\includegraphics*{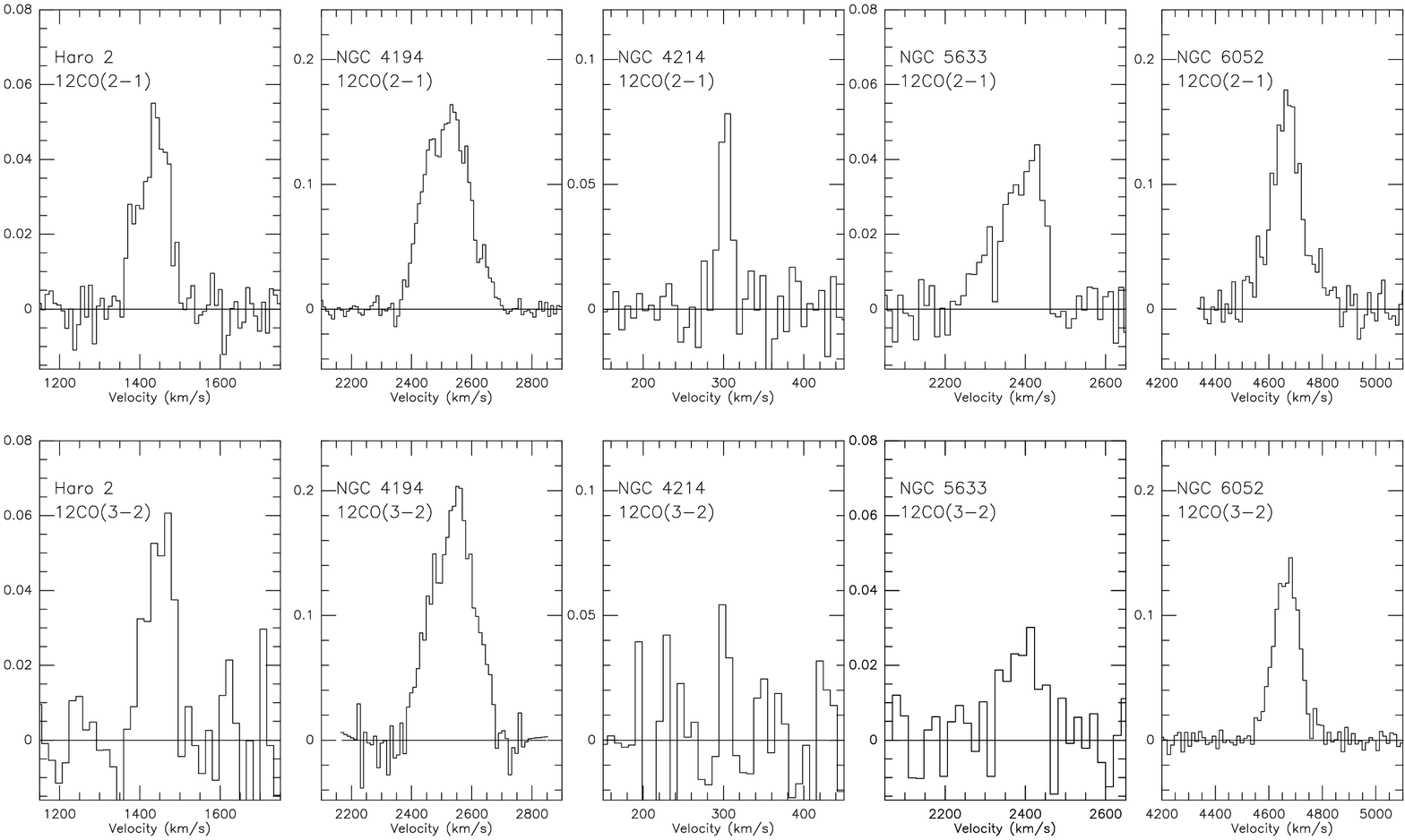}}}
\caption[]
{$J$=2-1 and $J$=3-2 $\co$ emission-line spectra observed towards the 
sample dwarf and peculiar galaxies. Horizontal scale is $V_{\rm LSR}$
in $\kms$, vertical scale is main-beam brightness temperature $T_{\rm mb}$
in Kelvins.
}
\label{spectra}
\end{figure*}

%Table 2
\begin{table*}
\caption[]{JCMT $\co$ observations of compact galaxies}
\begin{center}
\begin{tabular}{lrrcccrcrcc}
\hline
\noalign{\smallskip}
Galaxy   & RA(1950)   & DEC(1950) & $V_{\rm LSR}$ & Distance & $M_{BT}^{\circ}$ & Line &  Beam   & $T_{\rm mb}$ & $I_{\rm CO}$ \\
          &(h:m:s.s)  & (d:m:s)   & ($\kms$)      & (Mpc)    & (mag)            &      &  ($''$) & (mK)         & ($\kkms$)    \\  
\noalign{\smallskip}
\hline
\noalign{\smallskip}
NGC~55         & 00:12:30.5 & -39:28:55 &  +135 & 2.0 & -18.9 & J=2--1 & 21 &   75 &   2.5$\pm$0.2 \\
IC~10-SE       & 00:17:45.4 & +59:00:19 & --330 & 0.8 & -16.3 & J=2--1 & 21 &  735 &  13.0$\pm$0.3 \\  
               &            &           &       &     &       & J=3--2 & 14 & 1130 &  17.5$\pm$0.2 \\
               &            &           &       &     &       &        & 21 &  895 &  14.9$\pm$0.3 \\
               &            &           &       &     &       & J=4--3 & 11 &  490 &   9.4$\pm$0.9 \\
LGS~3          & 01:01:12.0 & +21:37:00 & --300 & 0.7 & -10.0 & J=2--1 & 21 &   11 &   1.4$\pm$0.3 \\
               &            &           &       &     &       & J=3--2 & 14 &   28 &   3.5$\pm$0.6 \\
2~Zw~40        & 05:53:05.0 & +03:23:07 &  +770 & 10  & -17:  & J=2--1 & 21 &    6 &   0.4$\pm$0.15 \\
               &            &           &       &     &       & J=3--2 & 14 & $<16$& $<0.8$\\
NGC~2537$^{a}$ & 08:09:42.8 & +46:08:33 &  +440 & 6.9 & -17.2 & J=2--1 & 21 &   23 &   1.1$\pm$0.2 \\
               &            &           &       &     &       & J=3--2 & 14 & $<11$& $<0.8$\\
NGC~2976       & 09:43:10.0 & +68:08:43 &   + 5 & 3.5 & -17.2 & J=2--1 & 21 &  103 &   4.1$\pm$0.2 \\
               &            &           &       &     &       & J=3--2 & 14 &   48 &   1.4$\pm$0.4 \\
Haro~2$^{b}$   & 10:29:22.7 & +54:39:24 & +1445 & 20  & -18.2 & J=2--1 & 21 &   52 &   4.5$\pm$0.2 \\
               &            &           &       &     &       & J=3--2 & 14 &   64 &   6.7$\pm$0.6 \\
               &            &           &       &     &       &        & 21 &   52 &   3.4$\pm$0.5 \\
NGC~4194$^{c}$ & 12:11:41.7 & +54:48:21 & +2506 & 39  & -20.1 & J=2--1 & 21 &  167 &  34.3$\pm$0.7 \\
               &            &           &       &     &       & J=3--2 & 14 &  213 &  39.3$\pm$2.5 \\
               &            &           &       &     &       &        & 21 &  177 &  26.7$\pm$2.6 \\
NGC~4214       & 12:13:11.2 & +36:35:47 & +280  & 4.1 & -17.9 & J=2--1 & 21 &   86 &   1.7$\pm$0.2 \\
               &            &           &       &     &       & J=3--2 & 14 &   78 &   1.2$\pm$0.45 \\
NGC~5633$^{d}$ & 14:25:37.2 & +46:22:33 & +2320 & 32  & -19.9 & J=2--1 & 21 &   36 &   5.2$\pm$0.7 \\
               &            &           &       &     &       & J=3--2 & 14 &   28 &   3.2$\pm$0.5 \\
NGC~6052$^{e}$ & 16:03:01.2 & +20:40:39 & +4725 & 65  & -20.7 & J=2--1 & 21 &  107 &  17.3$\pm$0.3 \\
               &            &           &       &     &       & J=3--2 & 14 &  151 &  18.0$\pm$0.8 \\
               &            &           &       &     &       &        & 21 &  117 &  12.9$\pm$0.8 \\   
\noalign{\smallskip}
\hline
\end{tabular}
\end{center}
Note: (a) Mkn~86 = Arp~6; (b) Mkn~33 = Arp~233; 
(c) Mkn~201 = 1~Zw~33 = Arp~160; (d) 1~Zw~89; 
(e) Mkn~297 = Arp~209.
\label{data}
\end{table*}

%Table 3
\begin{table*}
\caption[]{Summary of literature $\co$ data on observed galaxies}
\begin{center}
\begin{tabular}{lrcrclrcrclrcr}
\noalign{\smallskip}
\hline
\noalign{\smallskip}
J  &Beam&$I_{\rm CO}$ & Ref &\,\,\,\,& J &Beam&$I_{\rm CO}$ & Ref &\,\,\,\,& J &Beam&$I_{\rm CO}$ & Ref \\
   &($''$)& ($\kkms$) &  &        &   &($''$)& ($\kkms$) &  &        &   &($''$)& ($\kkms$) &\\  
\noalign{\smallskip}
\hline
\noalign{\smallskip}
&&{\bf NGC~55}         &  &&  && {\bf NGC~2976}&      && &&  {\bf NGC~4194}    &  \\
1-0& 43& 3.40$\pm$0.40 & 1&&1-0&100& 1.20$\pm$0.20& 3&&1-0&100& 1.55$\pm$0.19  & 3\\   
&&{\bf LGS~3}          &  &&1-0& 24& 4.12$\pm$0.17& 5&&1-0& 33& 17.0$\pm$2.0   &11\\
1-0& 45& 0.76$\pm$0.15 & 2&&2-1& 11& 1.98$\pm$0.30& 5&&1-0& 24& 29.3$\pm$0.26  & 5\\ 
&& {\bf 2~Zw~40}        &  &&  && {\bf Haro~2}     &  &&1-0& 22& 49.0$\pm$3.0   &12\\
1-0& 55& 0.75$\pm$0.35 & 3&&1-0& 55& 1.32$\pm$0.29& 1&&1-0& 15& 20.7$\pm$0.7   &13\\
1-0& 45& 0.62$\pm$0.15 & 2&&1-0& 55& 1.13$\pm$0.26& 3&&2-1& 22& 66.0$\pm$4.0   &12\\
1-0& 35& 1.10$\pm$0.50 & 4&&1-0& 22& 6.52$\pm$0.65& 6&&2-1& 11& 45.5$\pm$0.65  & 5\\   
1-0& 24& 0.23$\pm$0.05 & 5&&1-0& 22& 3.8$\pm$0.3  & 8&&3-2& 14& 17.7$\pm$0.4   &13\\
1-0& 22& 1.57$\pm$0.32 & 6&&2-1& 12& 6.2$\pm$0.4 & 8 && && {\bf NGC~4214-S}    &\\
2-1& 12& 0.88$\pm$0.16 & 6&&2-1& 12& 6.21$\pm$0.36 & 6&&1-0& 55& 0.65$\pm$0.10 &14\\
2-1& 11& 0.69$\pm$0.07 & 5&&3-2& 22& 2.3$\pm$0.3  & 7&&1-0& 55& 1.50$\pm$0.36  &15\\
3-2& 22& $<$0.9        & 7&&  && {\bf NGC~6052}   &  && &&  {\bf NGC~5633}     &\\
&&  {\bf NGC~2537}      &  &&1-0& 22&31.1$\pm$1.7  & 6&&1-0&100& 1.96$\pm$0.26  & 3\\
1-0& 24& 0.91$\pm$0.13 & 5&&1-0& 17&23.3$\pm$1.0  & 9&&&&&\\
1-0& 22& 1.16$\pm$0.38 & 6&&1-0& 15&30.6$\pm$8.0  &10&&&&&\\
2-1& 12& 1.12$\pm$0.17 & 6&&2-1& 12&23.8$\pm$0.7  & 6&&&&&\\				  
2-1& 11& 0.94$\pm$0.15 & 5&&3-2& 14&17.8$\pm$3.7  &10&&&&&\\
3-2& 22& $<$1.8        & 7&&   &   &              &  &&&&&\\
\noalign{\smallskip}
\hline
\end{tabular}		  
\end{center}
References: 1. Israel, Tacconi $\&$ Baas (1995); 2. Tacconi $\&$ Young
(1987); 3. Thronson $\&$ Bally (1987); 4. Gondhalekar et al. (1998);
5. Albrecht et al. (2004); 6. Sage et al. (1992); 7. Meier et al. 
(2001); 8. Barone et al. (2000); 9. Sofue et al. (1990); 10.  Yao
et al. (2003); 11. Aalto $\&$ H\"uttemeister (2000); 12. Casoli,
Dupraz $\&$ Combes (1992); 13. Devereux et al. (1994); 14. Taylor,
Kobulnicky $\&$ Skilman (1998); 15.  Thronson et al. (1988).
\label{litdata}
\end{table*}

\section{Observations}

\begin{figure*}[t]
\unitlength1cm
\begin{minipage}[]{18.0cm}
\resizebox{5.90cm}{!}{\rotatebox{270}{\includegraphics*{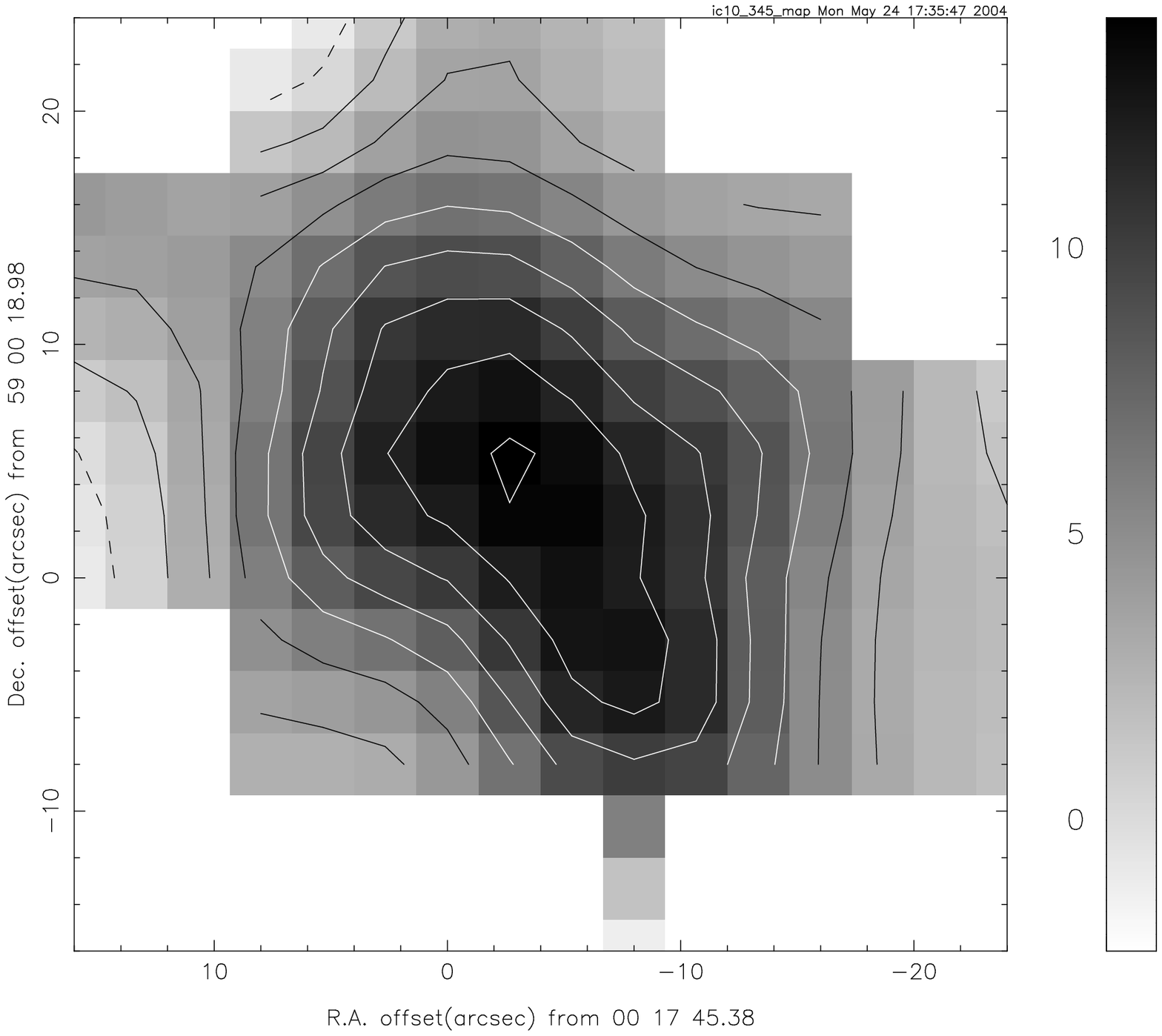}}}
\resizebox{7.38cm}{!}{\rotatebox{270}{\includegraphics*{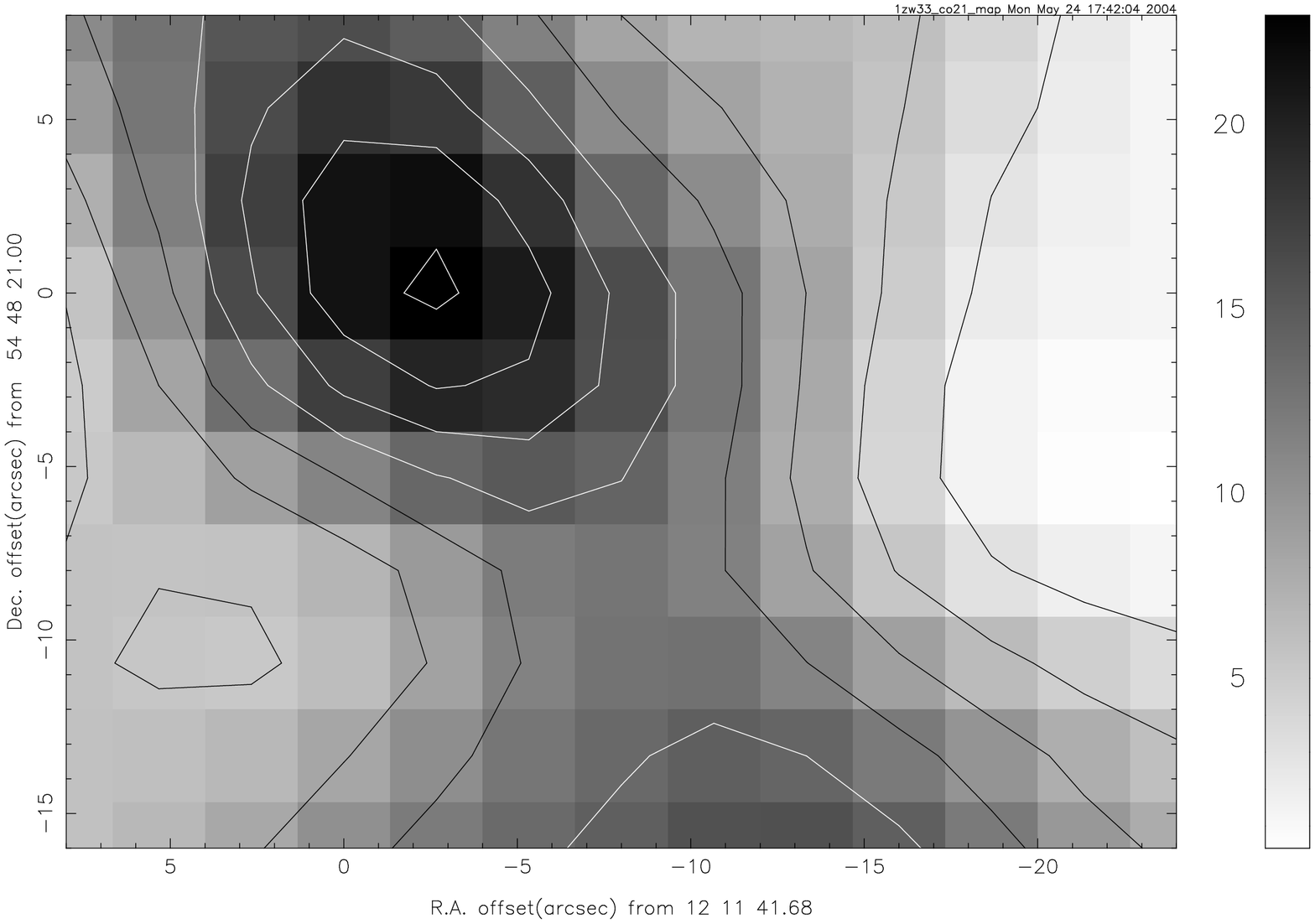}}}
\resizebox{4.36cm}{!}{\rotatebox{270}{\includegraphics*{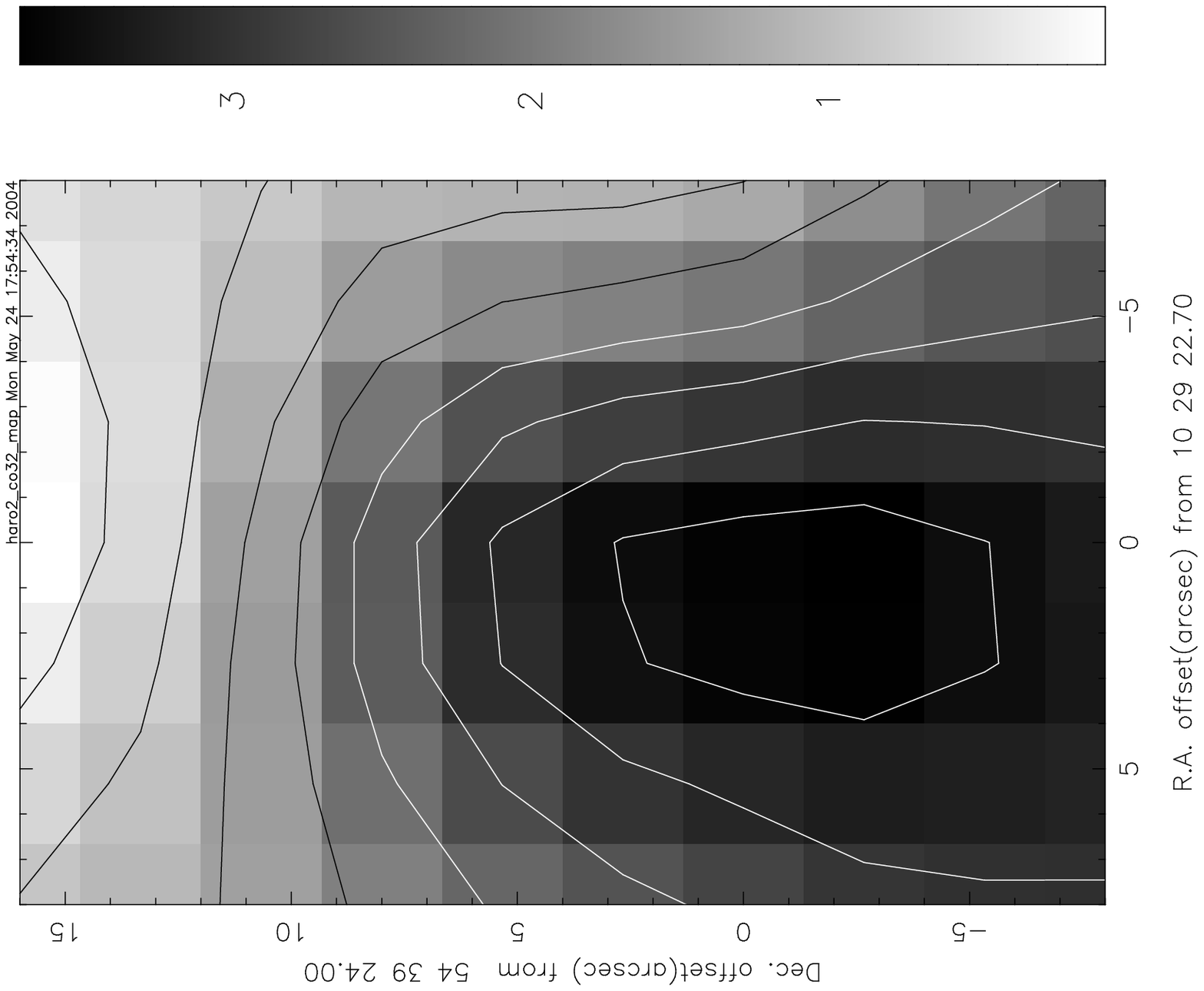}}}
\end{minipage}
\caption[] {Contour maps of integrated CO emission.  Lowest contour is
always at zero. Left: IC~10 ($J$=3--2) with main-beam brightness
temperature contours in multiples of 1 K. Center: NGC~4194 ($J$=2-1)
with contours in multiples of 4 K.  Right: Haro~2 ($J$=3-2) with
contours in multiples of 0.75 K.}
\label{maps}
\end{figure*}

All observations described in this paper were made with
the 15m James Clerk Maxwell Telescope (JCMT) on Mauna Kea (Hawaii)
\footnote{The James Clerk Maxwell Telescope is operated on a joint
basis between the United Kingdom Particle Physics and Astrophysics
Council (PPARC), the Netherlands Organisation for Scientific Research
(NWO) and the National Research Council of Canada (NRC).}, mostly
between July 1995 and December 1996. They were made in beamswitching
mode with a throw of $3'$ in azimuth using the DAS digital
autocorrelator system. When sufficient free baseline was available
(i.e when the detected line was sufficiently narrow), we subtracted
second or third order baselines from the profiles.  In all other
cases, linear baseline corrections were applied. All spectra were
scaled to a main-beam brightness temperature, $T_{\rm mb}$ = $T_{\rm
A}^{*}$/$\eta _{\rm mb}$. We used values of $\eta _{\rm mb}$
appropriate to the epoch of observation. These values were between
0.64 and 0.72 at 230 GHz, between 0.53 and 0.60 at 345 GHz, and
between 0.50 and 0.53 at 461 GHz. Spectra of observed positions are
shown in Fig.\,\ref{spectra}.  We also made small maps of the four
galaxies (IC~10, Haro~2, NGC~4194 and NGC~6052) distinguished by
relatively strong $J$=3--2 $\co$ line emission. These maps are shown
in Figs.\,\ref{maps}\,$\&$\,\ref{grid}.

%Figure2:Contour maps

\section{Results}

\subsection{Detections}

Details of the observed galaxies and the observational results
obtained are summarized in Table\,\ref{data}.  In this table, Column 1
gives the names, with alternative names in the notes at bottom.  Right
ascensions and declinations in columns 2 and 3 are those of the observed
(0,0) position. The receivers were tuned to the radial velocity given
in column 4; as is obvious from Fig.\,\ref{spectra} this was usually
very close to the CO velocity. Column 6 gives the integrated absolute
blue magnitude as retrieved from photometric data in the NASA-IPAC NED
and calculated for the assumed galaxy distance listed in Column 5.
there. Columns 7 and 8 identify the observed $\co$ transition and the
concomitant FWHM beamsize. Finally, columns 9 and 10 present the
observed peak main-beam brightness temperature, and the CO intensity
integrated over the observed profile. Quoted errrors are r.m.s. values;
upper limits are three times the r.m.s. value.

As Fig.\,\ref{spectra} and Table\,\ref{data} show, emission in the
$J$=2--1 $\co$ transition was detected from all observed galaxies,
although the emission from 2~Zw~40 was close to the detection
threshold.  2~Zw~40 was not detected in the $J$=3--2 transition nor
was NGC 2537 in spite of its clear detection in the $J$=2--1
transition.  By convolving the maps of IC~10, Haro~2, NGC~4194 and
NGC~6052 to a resolution of 21$''$, we also obtained for these
galaxies the $J$=3--2 $\co$ intensity matched to the $J$=2--1 beamsize
(Table\,\ref{data}).

\subsection{Line ratios}

%Table 4  Line Transition Ratios
\begin{table*}[t]
\begin{center}
\caption[]{Integrated rotational line ratios in compact galaxies}
\begin{tabular}{lcccc}
\hline
\noalign{\smallskip}
Galaxy & 12+lg[O]/[H] & $f_{60}/f_{100}$ & $\co$ (2--1)/(1--0) & $\co$ (3--2)/(2--1)) \\
       &              &                  & ($R_{21}$)          &   ($R_{32}$)\\
\noalign{\smallskip}
\hline
\noalign{\smallskip}
IC~10-SE    & 8.18 & 0.44 & 0.64$\pm$0.05 & 0.98$\pm$0.14  \\
LGS~3       & ---  & ---  & 0.4--1.8      & 1.1--2.5       \\ 
2~Zw~40     & 8.10 & 1.14 & 0.5:          & $<$1.5         \\
NGC~2537    & ---  & 0.54 & 0.93$\pm$0.14 & $<$0.8         \\
NGC~2976    & ---  & 0.35 & 0.96$\pm$0.07 & 0.72$\pm$0.11  \\
Haro~2      & 8.4  & 0.87 & 0.87$\pm$0.11 & 0.86$\pm$0.23  \\
NGC~4194    & ---  & 0.88 & 0.88$\pm$0.13 & 0.82$\pm$0.11  \\
NGC~4214-S  & 8.27 & 0.57 & 0.3--1.3      & 0.3--0.7       \\
NGC~5633    & ---  & 0.34 & 0.5--2.7      & 0.3--0.6       \\
NGC~6052    & ---  & 0.70 & 0.86$\pm$0.17 & 0.75$\pm$0.11  \\
\noalign{\smallskip}
\hline
\noalign{\smallskip}
NGC~3353    & 8.35 & 0.78 & \multicolumn{2}{c}{----- 0.56$\pm$0.12 -----} \\
\noalign{\smallskip}
\hline
\noalign{\smallskip}
He~2--10    & 8.4-8.9 & 0.91 & 0.97$\pm$0.16 & 1.16$\pm$0.23  \\
NGC~1569    & 8.19 & 0.90 & 1.15$\pm$0.10 & 1.05$\pm$0.09  \\
NGC~3077    & 9.02 & 0.59 & 0.83$\pm$0.16 & 1.12$\pm$0.22  \\
NGC~5253    & 8.15 & 1.05 & 1.43$\pm$0.29 & 1.06$\pm$0.25  \\ 
NGC~6822-HV & 8.21 & 0.55 & 1.28$\pm$0.30 & 1.00$\pm$0.20  \\
\noalign{\smallskip}
\hline
\label{ratio}
\end{tabular}
\end{center}
\end{table*}

We have determined rotational line ratios by directly comparing
transitions observed with beams of identical or very similar size,
using our own measurements and results from the literature, summarized
in Table\,\ref{litdata}. Where observations in identical beams were
lacking, we extrapolated CO fluxes to matching beamsizes.  This was
done either directly by plotting observed CO flux (Tables\,\ref{data}
$\&$ \ref{litdata}) as a function of beamsize and reading off the
desired flux value at the relevant beamsize, or indirectly by using
the equivalent CO source size.  The equivalent size can be determined
from observations with different beams (and not necessarily in the
same transition) under the assumption that the CO emission has a
radial gaussian brightness distribution with circular symmetry.  It
can also be estimated from high-resolution (array) maps of the CO
distribution.  Actual methods used are found in Sect.\,3.3 dealing
with the individual galaxies and the resulting ratios are given in
Table\,\ref{ratio}, together with metallicities [O]/[H] taken from
data compiled by Taylor et al. (1988) and Meier et al. (2001) and the
IRAS far-infrared 60$\mu$m/100$\mu$m flux-density ratios taken from
Lonsdale et al (1989).  Table\,\ref{ratio} consists of three parts:
the top section contains the galaxies observed by us, whereas the
remainder consists of data culled from the literature.  The bottom
section contains galaxies experiencing starburst activity.

Not included in Table\,\ref{ratio} is the dwarf galaxy UM~465 for
which CO measurements imply a (2--1)/(1--0) ratio of $1.16\pm0.29$
(Sage et al. 1992; Barone et al. 2000) and recently published results
by Albrecht et al. (2004) which allow direct determination of the
$2-1/1-0$ ratio for another six dwarf galaxies: NGC~145
(0.74$\pm$0.13); NGC~2730 (0.62$\pm$0.17); NGC~4532 ($0.59\pm0.12$);
NGC~6570 ($0.41\pm0.12$); NGC~7732 ($0.49\pm0.24$); IC~3521
($0.49\pm0.14$) as well as upper limits for four more: NGC~178 ($\leq
0.77$); NGC~1140 ($\leq 1.36$); NGC~3659 ($\leq 0.17$); NGC~4234
($\leq 0.80$).

On average, CO emission from compact galaxies is weak. As emission
from the $\13co$ isotope is usually at least an order of magnitude
less than that from $\co$, very few galaxies have been measured in
this line. In Table\,\,\ref{isotop} we list the available information
on the galaxies from Table\,\ref{ratio}. In effect, this sample is
limited to actively star-forming galaxies.
 
\subsection{Individual galaxies}

%Figure3 Grid map
\begin{figure}[h]
\unitlength1cm
\begin{minipage}[b]{9.cm}
\resizebox{9.cm}{!}{\rotatebox{270}{\includegraphics*{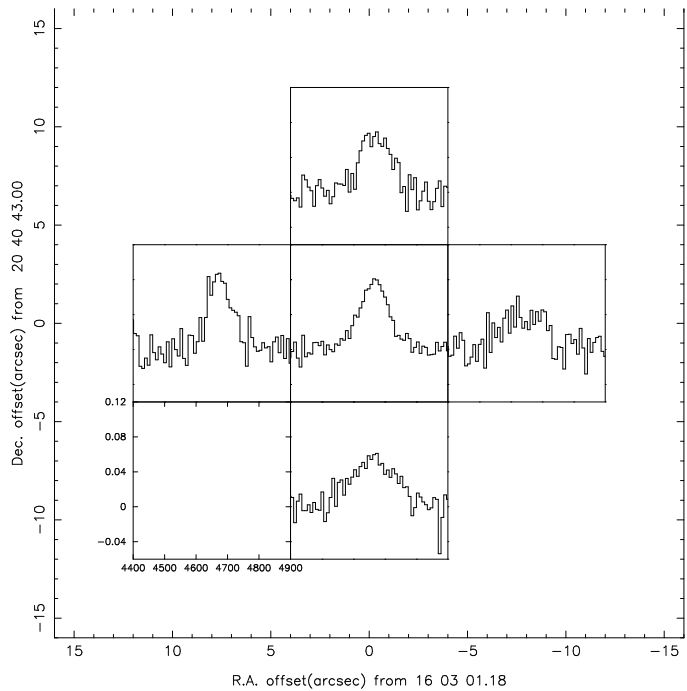}}}
\end{minipage}
\caption[] {Grid map of the $J$=3-2 CO emission from NGC~6052.
Horizontal scale is radial velocity $V_{\rm LSR}$ in $\kms$, vertical
scale is main-bean brightness temperature $T_{\rm mb}$ in Kelvins.}
\label{grid}
\end{figure}

{\bf NGC~55} is a small SB(s)m galaxy, roughly as we would expect to
see the LMC edge-on, member of the nearby Sculptor Group and rich in
interstellar gas. Dettmar $\&$ Heithausen (1989) used the SEST to make
a small $J$=1--0 $\co$ map of the brightest part.  At the position of
the CO peak in this map, Becker $\&$ Freundling (1991) also measured
the $J$=1--0 $\13co$ intensity. No further CO data on this interesting
galaxy have appeared in the literature.  Our $J$=2--1 spectrum (not
shown) was taken very close to the CO peak, at the same position as
our earlier $J$=1--0 spectrum (Israel, Tacconi $\&$ Baas, 1995).

{\bf IC~10} (dIrr IV) is an unusual Local Group galaxy; it is one of
the most luminous dwarf galaxies in the far-infrared (Melisse $\&$
Israel 1994), and appears to be the nearest representative of the
class commonly referred to as Blue Compact Galaxies (BCG). It contains
two major groupings of HII regions (see Hodge $\&$ Lee 1990; Yang $\&$
Skillman 1993; Chyzy et al. 2003), the brightest of which is
alternatingly known in the literature as IC~10A or IC~10-SE. The map
in Fig.\,\ref{maps} shows the CO distribution associated with
IC~10-SE, which resembles that of far-infrared and submillimeter
emission from heated dust (Thronson et al. 1990) and that of ionized
carbon ([CII]) emission (Madden et al. 1997). Studies of the dense
molecular medium in IC~10-SE were published by Petitpas $\&$ Wilson
(1998) and Bolatto et al. (2000). The (2--1)/(1--0) ratio in
Table\,\ref{ratio} was taken from the former; the (3--2)/(2--1) ratio
is the mean of their determination and our higher value.

{\bf NGC~1569} (IBm, Sbrst) is an extreme starburst dwarf galaxy
(Israel 1988). Weak CO detected by various authors (Greve et al. 1996;
Taylor et al. 1998, 1999; Meier et al. 2001; Albrecht et al. 2004) has
been mapped and discussed in M\"uhle's 2003 Ph.D. thesis from which we
took the ratios in Table\,\ref{ratio} (average of values from her
Table~3.9).  The extent of CO emission is slightly less than 30$''$.

{\bf LGS~3} (dIrr/dSph), a Local Group galaxy in Pisces, is one of the
faintest dwarf galaxies known. It was surprisingly detected in CO by
Tacconi $\&$ Young (1987) with a 45$''$ beam. Our measurements were
made at the same position, but with smaller beams. The range of ratios
in Table\,\ref{ratio} was obtained by assuming the CO source to be
pointlike or extended respectively.

{\bf 2~Zw~40} is a gas-rich galaxy alternately classified as a BCG or
an HII merger galaxy (Beck et al. 2002, and references therein).  CO
was detected by Thronson $\&$ Bally (1987), Tacconi $\&$ Young (1987),
and Sage et al. (1992). Our spectra were taken at the position used by
the latter. The integrated intensities as a function of beamsize
suggest an equivalent CO size of 23$''$ which was used to obtain the
(2--1)/(1--0) ratio in Table\,\ref{ratio}.

{\bf He~2-10} (I0pec, Sbrst) is a dwarf galaxy experiencing a strong 
starburst caused by an ongoing merger. The equivalent CO size is
about 12$''$, consistent with the OVRO $J$=1--0 CO map published by 
Kobulnicky et al. (1995). The CO ratios in Table\,\ref{ratio} were
taken from data in that paper, from Meier et al. (2001) and from Baas, 
Israel $\&$ Koornneef (1994).

{\bf NGC~2537} (SB(s)m pec) is also considered to be a BCG and often
referred to as Mkn~86 or, after its optical appearance, as the {\it
Bear Claw}. Not detected by Thronson $\&$ Bally (1987), $\co$ emission
was succesfully measured in the $J$=1--0 and $J$=2--1 transitions by
Sage et al. (1992). Our measurements were made at the same position,
and the line ratios follow directly from the data listed in
Tables\,\ref{data} $\&$ \ref{litdata}. NGC~2537 has also been mapped
in $J$=1--0 and $J$=2--1 $\co$ by Gil de Paz et al. (2002). Our $\co$
(2--1)/(1--0) ratio agrees with their less certain value 1.06$\pm$0.40

{\bf NGC~2976} (SAc pec, HII) is part of the M~81 group. Optically, it
is similar to NGC~2537 and NGC~1569 in having a complex central
structure surrounded by an extended and relatively featureless
distribution of stars. Bright HII regions occur at either end of the
major axis. CO was clearly detected by Thronson $\&$ Bally (1987) in a
large beam (100$''$). Our CO measurements refer to the center and do
not include either of the major HII region complexes. This central
part has also been mapped in $J$=1--0 $\co$ at high resolution by
Simon et al. (2003) using BIMA. Their map shows extended emission
consistent with the equivalent CO size of about 1$'$ suggested by the
single dish measurements. The values in Table\,\ref{ratio} were
calculated from the data in Tables\,\ref{data} $\&$ \ref{litdata} by
assuming a CO source size of 60$''$.

{\bf NGC~3077} (I0 pec, HII) is also a small member of the M~81 group.
CO emission from this galaxy was discovered by Becker et
al. (1989). Our measurements coincide with the peak in their
map. High-resolution OVRO maps of the galaxy in the $J$=1--0 and
$J$=2--1 CO transitions were published by Meier et al. (2001) and
Walter et al. (2002) and show significant substructure.  The ratios in
Table\,\ref{ratio} were derived from data by Becker et al. (1989),
Meier et al. (2001) and Albrecht et al. (2004) for an equivalent CO
source size of $25''$.

{\bf Haro 2} (Im pec, HII) is another BCG also known as Mkn~33. CO was
detected by Thronson $\&$ Bally (1987), Sage et al. (1992), Israel,
Tacconi $\&$ Baas (1995), Barone et al. (2000), and Meier et al.
(2001) including measurement of the $J$=2--1 and $J$=3--2 $\co$
transitions.  An OVRO map of the $J$=1--0 $\co$ distribution was
presented by Bravo-Alfaro et al. (2004) and shows that most of the
emission originates in a source of size 15$''$-20$''$
(cf. Fig.\,\ref{maps}). Our (2--1)/(1--0) $\co$ ratio is based on the
mean of the $J$=1--0 CO fluxes in Table\,\ref{litdata}. The implied
(3--2)/(2--1) ratio appears to be dependent on the beamsize used.
The mean ratio for the 21/22$''$ beam is 0.63; whereas the the
mean for the 12/14$''$ beam is 1.08. In Table\,\ref{ratio} we
list the average of these two.

{\bf NGC~3353 = Haro~3 = Mkn~35} (Irr, HII) is a BCG detected by
Thronson $\&$ Bally (1987), Tacconi $\&$ Young (1987), Sage et
al. (1992), and Meier et al. (2001).  The available data only allow a
reliable determination of the (3--2)/(1--0) ratio, as the $J$=2--1
measurement by Sage it al. (1992) appears to be too low.

{\bf NGC~4194} (IBm pec, SB0 pec, HII) is also known as the BCG
1~Zw~33, Mkn~201 or the {\it Medusa Merger}. Measurements of the lower
three $\co$ transitions have been published by Casoli et al. (1992),
Devereux et al. (1994) and Aalto et al. (2001).  An OVRO $J$=1--0
$\co$ map was presented by Aalto $\&$ H\"uttemeister (2000). The
equivalent CO size of 8$''$ suggested by the single-dish measurements
is consistent with the extent of the major CO concentration in their
map. The data summarized in Table\,\ref{litdata} are, unfortunately,
not entirely consistent. In particular, we have decided to ignore the
high $J$=2--1 $\co$ value from Casoli et al. (1992), which is
inconsistent with all other measurements. To obtain the
$J$=2--1/$J$=1--0 ratio, we have divided our $J$=2--1 value from
Table\,\ref{data} by the mean of the 22/24$''$ $J$=1--0 data in
Table\,\ref{litdata}.  For the $J$=3--2/$J$=2--1 ratio we have taken
the mean of our own (21$''$) values in Table\,\ref{data} (0.78) and
the 11/14$''$ values in that Table and in Table\,\ref{litdata}
(0.86). In view of the uncertainties, we have not attempted to correct
the latter for finite source and beam sizes.

%Table 5  Isotopic Line Ratios
\begin{table}[]
\begin{flushleft}
\caption[]{Integrated isotopic line ratios in compact galaxies}
\begin{tabular}{lcccc}
\hline
\noalign{\smallskip}
Galaxy      &\multicolumn{3}{c}{$\co/\13co$ }& Ref \\   
            & $J$=1--0 & $J$=2--1 & $J$=3--2 & \\
            &($r_{1}$) &($r_{2}$) &($r_{3}$) & \\    
\noalign{\smallskip}
\hline
\noalign{\smallskip}
NGC~55      & 15$\pm$4 &   ---    &    ---   & 1 \\ 
IC~10-SE    &  9$\pm$1 & 13$\pm$3 & 12$\pm$2 & 2 \\
He~2--10    & 20$\pm$4 & 20$\pm$4 & 16$\pm$5 & 3 \\
NGC~1569    & 23$\pm$8 & 36$\pm$9 &    ---   & 4 \\
NGC~4194    & 19$\pm$4 &   ---    &    ---   & 5 \\
NGC~6822-HV & 23$\pm$7 &   ---    & 11$\pm$3 & 6 \\
\noalign{\smallskip}
\hline
\label{isotop}
\end{tabular}
References: 1. Becker $\&$ Freundling (1991); 2. Petitpas $\&$ Wilson
(1998); Bolatto et al. (2000); redetermined from archive data; 3.
Baas et al. (1994); Kobulnicky et al. (1995); 4. M\"uhle (2003);
5. Aalto et al. (2001); 6. Israel et al. (2003), redetermined.
\end{flushleft}
\end{table}

{\bf NGC~4214} (IAB(s)m, HII) was measured in $\co$ by Tacconi $\&$
Young (1985), Thronson et al. (1988), Ohta et al. (1993), Becker et
al. (1995), and Taylor et al.  (1998). An OVRO map by Walter et
al. (2000) shows three CO concentrations along the major axis
separated by 40$''$, each about 10--15$''$ in size. Our measurements
(cf. Fig\,\ref{maps}) sample the southern maximum. The range of ratios
in Table\,\ref{ratio} refer to source sizes varying from great extent
(high ratio) to essentially unresolved (low ratio).  The (3--2)/(2--1)
ratio cannot be determined accurately, but the upper limit is
physically significant.

{\bf NGC~5253} (Impec, HII, Sbrst), also known as Haro~10, is a small
but pronounced starburst member of the M~83 group. The $J$=1--0
observations in various beams (Wiklind $\&$ Henkel 1989; Turner et
al. 1997; Taylor et al. 1998) suggest an equivalent CO size of
35$''$. This, together with the $J$=2--1 and $J$=3--2 obervations by
Meier et al.  (2001) implies the ratios given in Table\,\ref{ratio}.

{\bf NGC~5633} ((R)SA(rs)b) was detected in CO by Thronson $\&$ Bally
(1987). No other CO measurements have appeared in the literature, so
that we can only constrain possible ratios by the assumption of either
pointlike or extended CO emission. Again, only the (3--2)/(2--1)
ratio is sufficiently constrained to have physical meaning.

{\bf NGC~6052} is an otherwise unclassified BCG more commonly known as
Mkn~297 with an optical size of about 20$''$, possibly the result of a
merger. Comparison of all available measurements suggests that Sage et
al.  (1992) overestimated the $J$=1--0 intensity; the ratio derived
from 14$''$ beam observations (assuming identical source structure in
all CO transitions) is more plausible and given in Table\,\ref{ratio}.
Our $J$=3--2 map (Fig.\,\ref{grid}) shows the CO source to be resolved
whereas the $J$=1--0 map by Sofue et al. (1990) only shows marginal
resolution.  Taken together, these results suggest a CO extent of
about 15$''$.

{\bf NGC6822} is a Local Group galaxy similar to the LMC. The entries
in Table\,\ref{ratio} were taken from Israel et al. (2003) and refer
to individual pointings on the CO cloud complex associated with the
star-formation region Hubble~V.

\section{Analysis and discussion}

\subsection{Physical condition of the gas}

%Table 6
\begin{table*}
\caption[]{Physical parameters of model clouds}
\begin{center}
\begin{tabular}{ccccccccccccccc}
\hline
\noalign{\smallskip} 
& Kin. & Gas & Column &Contribution &\multicolumn{5}{c}{Observed}&\multicolumn{5}{c}{Model}\\
& Temp.& Density&Density&to $J$=2--1&\multicolumn{5}{c}{Ratios}  &\multicolumn{5}{c}{Ratios}\\
&$T_{\rm k}$&$n(\h2$)&$N$(CO)/d$V$&Emission&$R_{21}$&$R_{32}$&$r_{1}$&$r_{2}$&$r_{3}$&$R_{21}$&$R_{32}$&$r_{1}$&$r_{2}$&$r_{3}$\\
&(K) &($\cc$)&($\cm2/\kms$) &&&&&&&&&&&\\
\noalign{\smallskip}
\hline
\noalign{\medskip}
\multicolumn{15}{c}{NGC~2537, NGC~2976, Haro~2, NGC~4194, NGC~6052} \\
\noalign{\smallskip}
& 60 &3000&$0.6-1.0\times10^{17}$&1.00&0.91&0.80&15&--&--&0.93&0.78&16& 9&14\\
\noalign{\smallskip}
\hline
\noalign{\medskip}
\multicolumn{15}{c}{IC~10-SE  }\\
\noalign{\smallskip}
1 &  30  &$10^{4}-10^{5}$ &$1.0\times10^{17}$ &0.15 &0.64&0.98& 9&13&12&0.76&0.93&11&11&13\\
2 & 100  & 100     &$1.0\times10^{17}$ &0.85 &    &    &  &  &  &    &    &  &  &  \\
\noalign{\medskip}
\multicolumn{15}{c}{NGC~1569}\\
\noalign{\smallskip}
1 &  100 &$10^{5}$ &$1.0\times10^{17}$ &0.25 &1.15&1.05&23&36&--&1.23&0.99&23&20&18\\
2 &  100 & 1000    &$0.3\times10^{17}$ &0.75 &    &    &  &  &  &    &    &  &  &  \\
\noalign{\medskip}
\multicolumn{15}{c}{He~2-10, NGC~3077}\\
\noalign{\smallskip}
1 &  60  &$10^{5}$ &$0.6\times10^{17}$ &0.20 &0.90&1.14&20&20&16&0.99&0.83&21&18&17 \\
2 &100$\pm$50&500--1000&$0.3\times10^{17}$&0.80&  &    &  &  &  &    &    &  &  &   \\
\noalign{\medskip}
\multicolumn{15}{c}{NGC6822-HV, NGC~5253}\\
\noalign{\smallskip}
1 &  30 &$10^{5}$  &$0.6\times10^{17}$ &0.75 &1.35&1.03&23&--&11&1.45&0.99&22&14&11\\
2 &30-150&$10^{5}$ &$0.3\times10^{17}$ &0.25 &    &    &  &  &  &    &    &  &  & \\
\noalign{\smallskip}
\hline
\label{model}
\end{tabular}
\end{center}
\end{table*}

The observed $\co$ and $\13co$ transitions can be analyzed to provide
constraints on the physical condition of the molecular gas in the
compact galaxies concerned.  To this purpose we have used the
large-velocity gradient (LVG) radiative transfer models described by
Jansen (1995) and Jansen et al. (1994).  They provide model line
intensities as a function of three input parameters: gas kinetic
temperature $T_{\rm k}$, molecular hydrogen density $n(H_{2})$ and CO
column density per unit velocity ($N({\rm CO})$/d$V$).  Comparison of
model to observed {\it line ratios} allows identification of the
physical parameters best describing the actual conditions in the
observed source.  Beam-averaged properties are determined by comparing
observed and model line intensities.  Tables\,\ref{ratio} and
\ref{isotop} show that in several galaxies the input parameters are
not fully constrained by the observations.  At least some constraint
on gas conditions can be derived for NGC~2976, NGC~3077, Haro~2
(Mkn~33), NGC~4194 (Mkn~201), NGC~5253, and NGC~6052 (Mkn~297).
Sufficient information for a proper analysis is provided only by (the
star-forming complexes in) IC~10-SE, NGC~1569, NGC~6822-HV and
He~2-10.

\subsubsection{Limitations of single-component modelling}

First, we have attempted to fit the observed line ratios with a
molecular gas at a single temperature, a single density and a single
velocity gradient by searching a grid of model intensity ratios
corresponding to temperatures $T_{\rm k} = 10-250$ K, densities
$n(\h2) = 10^{2}-10^{5} \cc$, and gradients $N(CO)/{\rm d}V =
6\times10^{15}-3\times10^{18} \cm2/\kms$) for values matching the
observed intensity ratios.  Among the galaxies listed in
Table\,\ref{ratio}, five (NGC~2537, NGC~2976, Haro~2, NGC~4194 and
NGC~6052) have almost identical line ratios with mean values
(2--1)/(1--0) = 0.90 and (3--2)/(2--1) = 0.78. These ratios are
accurately reproduced ({Table\,\ref{model}) by a warm and moderately
dense molecular gas of temperature $T_{\rm kin} = 60(+40,-20)$ K,
density $n(\h2) = 3000 \cc$, and gradient $N(\rm CO)/{\rm d}V =
0.6-1.0\times10^{17} \cm2 (\kms)^{-1}$. The $J$=1--0 $\co/\13co$ ratio
in NGC~4194 (Table\,\ref{isotop}) agrees well with this. The
intensities imply a filling factor $3.3\times10^{-3}$ and {\it
beam-averaged} CO column-densities range from $\approx
1\times10^{16}\,\cm2$ (NGC~2976, Haro~2) to $5\times10^{16}\,\cm2$
(NGC~6052) to $1\times10^{17}\,\cm2$ (NGC~4194). The relatively low
(2--1)/(1--0) ratios (mean: 0.56) for the Albrecht et al. (2004)
galaxies (Sect.\,3.2), and for 2~Zw~40 (Table\,\ref{ratio}), suggest
the dominating presence of fairly low-density ($n(\h2) \leq 800 \cc$,
$N(\rm CO)/{\rm d}V = 0.3-0.6\times10^{17} \cm2 (\kms)^{-1}$r gas at
undetermined temperatures.

The physical meaning of these results should be established by further
observations, especially of the $\13co$ isotope, but we suspect that
additional information will only serve to rule out single-component
fits after all.  It is perhaps telling that no other galaxy in
Table\,\ref{ratio} is satisfactorily fitted by a single component.
Moreover, even in cases where a single component appears to provide a
good (LVG) fit to the observations, caution should be exercised in
accepting the result as a physical reality, especially when the fit is
based on limited data. This may be illustrated by the case of
IC~10-SE.  The relative intensities of the $J$=1--0, $J$=2--1 and
$J$=3--2 $\co$ lines together with the $J$=1--0 and $J$=2--1
$\co/\13co$ isotopic ratios are perfectly fitted by a single hot
($T_{\rm kin} = 100$ K) and tenuous ($n(\h2) = 100 \cc$) molecular gas
component.  Note that this result is based on a total of four line
ratios.  As many determinations in the literature are based on three
or only two ratios, this is more than the number commonly used in such
analyses.  Yet, this apparently excellent fit completely fails to
correctly predict the other two ratios measured for IC~10-SE.  The
observed modest $J$=3-2 $\co/\13co$ ratio of eleven is four times
lower than predicted, and the modelled $J$=4--3 $\co$ intensity falls
short of the observed value by a factor of two or more.

Although this might suggest that single-component LVG analysis is
physically irrelevant, this is not quite true.  The kinetic
temperatures and spatial densities implied by such fits often actually
occur in the source. However, when they do, temperature and density
generally {\it do not refer to the same volume of gas}, as will be
clear from the following.

\subsubsection{Dual-component modelling}

For almost all of the sample galaxies, good fits based on two gas
components could be obtained.  Although the number of free parameters
exceeds the number of independent measurements, this is a meaningful
result because the actual combination of physically allowed parameters
is sufficiently constrained.  In order to reduce the number of free
parameters, we have assumed identical CO isotopical abundances for
both gas components and assign the specific value [$\co$]/[$\13co$] =
40.  Different choices are possible, but reasonably small changes, for
instance to ratios of 50 or 60 rarely lead to very different outcomes.
We identified acceptable fits by searching a grid of model parameter
combinations (covering the same range of temperatures, densities and
gradients as described above) for matching model and observed line
ratios, for various relative contributions of the two components.

We have rejected all solutions in which the denser gas component is
also hotter than the more tenuous component, as we consider this
physically unlikely on the large linear scales observed. From the
remainder of solutions, we have selected characteristic examples and
listed these in Table\,\ref{model}. The solutions are not unique, but
delineate a range of values in particular parameter space
regions. Variations in input parameters may compensate for one
another, causing somewhat different input combination to yield
identical line ratios, as indicated in the Table entries.

%Table 7
\begin{table*}
\caption[]{Beam-averaged physical parameters}
\begin{center}
\begin{tabular}{lcccccr}
\hline
\noalign{\smallskip} 
Galaxy & \multicolumn{2}{c}{Abundances}    &\multicolumn{2}{c}{Column Density} & Mass          &CO-$\h2$ Ratio\\
& $N_{\rm C}/N$(CO)  &$N_{\rm H}/N_{\rm C}$& $N({\rm HI})$ & $N(\h2)$  &$M(\h2)$       &$X$\\
&                    &                     &($\cm2$)&($\cm2$) &($\Msun$)&($\cm2/\kkms$)\\
\noalign{\smallskip}
\hline
\noalign{\smallskip}
Haro 2     &21$\pm$2 &$4\times10^{4}$&$1.6\times10^{21}$&$ 5\times10^{21}$&$7\times10^{7}$&$ 9\times10^{20}$\\
IC~10-SE   & 5$\pm$1 &$1\times10^{5}$&$3.6\times10^{21}$&$24\times10^{21}$&$3\times10^{6}$&$12\times10^{20}$\\
NGC~1569   &46$\pm$23&$1\times10^{5}$&$3.7\times10^{21}$&$11\times10^{21}$&$1\times10^{7}$&$72\times10^{20}$\\
He~2-10    &11$\pm$2 & ---           &$1.9\times10^{21}$& ---             & ---           &$<4\times10^{20}$\\
NGC~6822-HV&22$\pm$3 &$6\times10^{4}$&$1.6\times10^{21}$&$10\times10^{21}$&$5\times10^{5}$&$60\times10^{20}$\\
\noalign{\smallskip}
\hline
\label{beamave}
\end{tabular}
\end{center}
Note: HI column densities from Bravo-Alfaro et al. (2004: Haro~2), Madden 
et al. (1997: IC~10), Stil $\&$ Israel (2002: NGC~1569), Kobulnicky et al. 
(1995: He~2-10), Israel et al. (2003: NGC~6822).
\end{table*}

%Table 8  Comparisons
\begin{table*}
\begin{center}
\caption[]{Environment and molecular gas parameters}
\begin{tabular}{lccccccccc}
\hline
\noalign{\smallskip}
Galaxy & \multicolumn{3}{c}{$\co$} & \multicolumn{3}{c}{$\co/\13co$} & $N_{\rm C}/N({\rm CO})$ & log $N({\rm CO})$ & log $N(\h2)$ \\ 
& (2--1)/(1--0) & (3--2)/(2--1) & (3--2)/(1-0) & (1--0) & (2--1) & (3-2) & & ($\cm2$) & ($\cm2$) \\
\noalign{\smallskip}
\hline
\noalign{\smallskip}
\multicolumn{10}{c}{Compact Galaxies} \\
\noalign{\smallskip}
\hline
\noalign{\smallskip}
Type I (Starburst) & 1.1 & 1.1 & 1.2 & 20  & 23: & 13  & 21  & 16.6 & 22.2 \\
Type II (Active)   & 0.9 & 0.8 & 0.7 & 17: & --- & --- & 21  & 16.2 & 21.7 \\
Type III (Quiet)   & 0.5 & --- & --- & --- & --- & --- & --- & ---  & ---  \\
\noalign{\smallskip}
\hline
\noalign{\smallskip}
\multicolumn{10}{c}{Spiral Galaxy Centers}\\
\noalign{\smallskip}
\hline
\noalign{\smallskip}
Starburst          & 1.0 & 0.7 & 0.7 & 10  & 10  & 11  &  3 & 18.1  & 21.7 \\
Quiet Nucleus      & 0.5 & 0.6 & 0.3 &  7  &  6  & --  &  6 & 18.0  & 21.6 \\
\noalign{\smallskip}
\hline
\noalign{\smallskip}
\multicolumn{10}{c}{LMC/SMC Clouds}\\
\noalign{\smallskip}
\hline
\noalign{\smallskip}
Star-Forming       & 1.1 & 0.9 & 1.0 & 12  &  8  & --  &  - & --    & --   \\
Quiet              & 0.7 & 0.5 & 0.4 & 10  & 14  & --  &  - & --    & --   \\
\noalign{\smallskip}
\hline
\label{compa}
\end{tabular}
\end{center}
\end{table*}

The observed line ratios for the star-forming regions in all six
galaxies listed in Table~5 require the presence of a rather dense
component (typically $n_{\h2} = 10^{5} \cc$) but these components vary
from fairly cool (IC~10-SE, NGC~6822-HV and probably NGC~5253) to warm
(He2-10 and probably NGC~3077) to hot (NGC~1569). Column densities are
well-established in the range $N({\rm CO})$/d$V$ = 0.6--1.0 $\times
10^{17} \cm2/\kms$. This dense component is associated with a second
component that is generally less dense, but hotter (typically 100 K).
The actual density of this second component ranges from $n_{\h2} = 100
\cc$ (IC~10-SE) to $n_{\h2} \approx 10^{3} \cc$ (NGC~1569, He~2-10 and
probably NGC~3077). No signature of low-density gas is apparent in
NGC~6822-HV and possibly NGC~5253. In these objects, the hot second
component is still very dense (typically $10^{5} \cc$). Note that the
notable physical difference between e.g. NGC~1569 and NGC~6822 is not
immediately obvious as these objects have very similar $\co$ line
ratios and identical $J$=1--0 $\co/\13co$ ratios. The important
discriminators are the remaining $\co/\13co$ ratios. A relatively low
$J$=3-2 $\co/\13co$ ratio (NGC~6822) is incompatible with a
significant presence of low-density gas at any temperature, whereas a
high $J$=2-1 $\co/\13co$ (NGC~1569) requires the presence of very hot
and very dense gas, or just low-density gas, or a combination of both.
Even then, we find it difficult (as Table\,\ref{model} shows) to
reproduce the very high $J$=2-1 $\co/\13co$ apparently appropriate to
NGC~1569.  This suggests that the observations are in error, or that
our assumptions are incorrect. Neither possibility can be ruled out.
NGC~1569 is one of the most extreme (post) starburst known (cf. Israel
1988) and its ISM appears to be subjected to intense processing
(Lisenfeld et al. 2002). Thus, the less abundant $\13co$ isotope may
have selectively destroyed, so that its actual abundance is much less
than assumed in our modelling. This would indeed lead to higher
$\co/\13co$ ratios. In the absence of further data we have chosen not
to pursue this possibility, but it does suggest that (re)determination
of the $\13co$ intensities in NGC~1569 might be fruitful.

\subsection{Beam-averaged molecular gas properties}

The fraction of carbon locked up in the carbon monoxide molecule
(i.e. the $N({\rm CO})/N_{\rm C}$ ratio) as given by the chemical
models by Van Dishoeck $\&$ Black (1988) is strongly dependent on the
total carbon (and molecular hydrogen) column density. At column
densities $N_{\rm C} > 10^{18}\,\cm2$ practically all gas-phase carbon
is in CO, whereas at column densities $N_{\rm C} < 10^{17}\,\cm2$
essentially all carbon is in atomic form. Using these chemical models
and the radiative transfer model parameters summarized in
Table\,\ref{model}, we have estimated the {\it beam-averaged} carbon
monoxide column densities as well as the {\it beam-averaged} column
densities of all (atomic and molecular) carbon present in the
gas-phase. Total carbon column densities $N_{\rm C}$ may be converted
into total hydrogen column densities $N_{\rm H}$ as a function of the
[C]/[H] abundance and the fraction of carbon in the gas-phase.  For
galaxies with metallicities below solar, the [C]/[H] abundance can be
estimated from the [C]/[O] versus [O]/[H] diagrams given by Garnett et
al (1999) and the [O]/[H] metallicities summarized in
Table\,\ref{ratio}. We furthermore assume a gas-phase carbon-depletion
factor $\delta_{\rm C}$ = 0.27. By subtracting observed neutral
hydrogen column densities $N(\rm HI)$ from the total hydrogen column
densities thus obtained, beam-averaged molecular hydrogen column
densities are determined. Finally, these are used to obtain total
molecular hydrogen masses and estimates of the CO-to-$\h2$ conversion
factor $X$. Table\,\ref{beamave} gives the results for those galaxies
where this procedure was meaningful. We find that at least in the
low-metallicity starburst objects, overall CO column densities are low
(which is reflected by the relatively low velocity-integrated
intensities in Table\,\ref{data}). We also find that in these objects,
only a small fraction of all carbon is in CO and that most carbon
should be in neutral or ionized atomic form. As a consequence,
molecular hydrogen densities are not proportionally lower. Low
integrated CO intensities and relatively normal molecular hydrogen
column densities together imply the high $X$-values given in
Table\,\ref{beamave}.

\subsection{Molecular gas in compact galaxies}

Although the database is admittedly still limited, we can make a few
statements as to the general nature of molecular gas in compact and
dwarf irregular galaxies. On the basis of CO spectroscopy we may
distinguish different types of environment in compact galaxies, with
properties summarized in Table\,\ref{compa}. We have included results
obtained in a similar way for a sample of high-metallicity starburst
galaxy centers (NGC~253, IC~342, Maffei~2, M~83, and NGC~6946 --
Israel et al. 1995, Israel $\&$ Baas 2001, 2003) and for a quiescent
centers such as occur in NGC~7331 (Israel $\&$ Baas 1999) and the
Milky Way (Bennett et al. 1994). The former are characterized by CO
intensities slowly decreasing with increasing rotational transition,
as well as $\co/\13co$ ratios of about 10.  The latter are weaker CO
emitters, more difficult to detect and less frequently included in
surveys. They have CO intensities much more rapidly decreasing with
rotational transition, and they have lower $\co/\13co$ ratios around 6
-- 7. Finally, we also included average ratios obtained for both
quiescent and actively star-forming molecular clouds in the
low-metallicity environments of the Large and the Small Magellanic
Clouds (Bolatto et al. 2000, 2005).

{\it Type I (starburst compact)} has CO emission with $\int{T_{\rm
mb}{\rm d}V}$ rising with increasing $J$-transition: both the
(2--1)/1--0) and the (3--2)/(2--1) ratios are close to or even
exceeding unity. At the same time, the $\co/\13co$ isotopical ratios
are quite high, typically of the order of 20. The CO results obtained
towards NGC~1569, He~2-10, NGC~5253 and NGC~6822-HV are all in this
category, as are those of the object UM~465 mentioned earlier. The
molecular clouds sampled are dominated by intense star-formation
activity. This may be representative of a large part of the galaxy
(NGC~1569, He~2-10) or only of a specific location (NGC~6822-HV). The
rotational line ratios imply high overall temperatures, and the
isotopical ratios are characteristic of relatively low molecular gas
beam-averaged column densities. This is clearly illustrated in
Table\,\ref{compa}.  The line ratios of Type I compact galaxies are
very similar to those of spiral galaxy starburst centers and the
low-metallicity starforming clouds in LMC and SMC, except for much
higher $\co/\13co$ ratios. This is characteristic of similar (high)
temperatures but lower optical depths in the starburst dwarf galaxies.
Hence, in the compact galaxies beam-averaged CO column densities are
more than an order of magnitude lower than those in starburst galaxy
centers and star-forming LMC/SMC clouds.  In the compact galaxies most
carbon is in atomic form. The beam-averaged molecular hydrogen columns
are very similar to those of the starburst centers.  The line ratios 
of IC~10-SE are different from any other set. They most resemble a
mixture of active and cold LMC/SMC clouds. 

Because integrated CO intensities are low and the $\h2$ column
densities implied by the analysis are not, $X$-factors in compact
galaxies very significantly exceed those of starburst centers.  We
have calculated values $X = (12, 72, 60) \times 10^{20}
\cm2(\kkms)^{-1}$ for IC~10, NGC~1569 and NGC~6822 respectively. These
values are in good agreement with conclusions reached earlier and
independently for IC~10 by Madden et al. (1997) and for NGC~1569 by
Lisenfeld et al. (2002). They also lend further support to the strong
metallicity dependence of the $X$-factor proposed by Israel (1997,
2000).  

{\it Type II (active compact)} has Haro~2 as a prototype. The line
parameters are similar to those of Type I objects, but the rotational
line ratios are lower and less than unity. The difference is clearly
seen in the $\co$ (3--2)/(1--0) ratio which is almost half that of the
Type I average.  Most of the galaxies in the top panel of
Table\,\ref{ratio}, including NGC~3353 (Haro~3) are of Type II.  They
are actively forming stars, but the CO spectroscopy is not dominated
by star-forming regions as in the Type I objects. IC~10 and NGC~6822,
now included in Type I, would probably fall in this category if we had
sampled the whole galaxy instead of only the most actively
star-forming part.

Even for Haro~2 the observational data are insufficient to allow a
meaningful two-component analysis, so that the derived parameters are
based only on an overly simplified single-component analysis.
Nevertheless, as in Type I objects, carbon monoxide appears to be only
$\approx 5\%$ of all gas-phase carbon. Carbon monoxide columns appear
two orders of magnitude lower than those in starburst nuclei, but
molecular hydrogen columns are virtually identical. For Haro~2 we find
$X = 9 \times 10^{20} \cm2(\kkms)^{-1}$ practically identical to the
value derived in a different manner by Barone et al. (2000). Although
this is much lower than the average derived for the Type I objects, it
is very close to the value expected for this metallicity in the
diagrams by Israel (1997, 2000). If we assume, just for the sake of
argument, the same metallicity for all type II galaxies, we can
perform an analysis similar to that of Haro~2 also for the other
galaxies in this particular subsample. Not surprisingly, we obtain
very similar CO, C and $\h2$ columns (with a mean log $N(\h2)$ =
21.8). For NGC~2537 and NGC~2976 we would find $X = 10-30 \times
10^{20} \cm2/\kkms$, but for the much more distant galaxies NGC~4194
and NGC~6052 we would obtain $X$ values an order of magnitude lower,
not substantially different from the Solar Neighborhood value $X = 2
\times 10^{20} \cm2/\kkms$. The fact that these are the most luminous
galaxies in the sample (cf. Table\,\ref{data}) is probably more
relevant to this result than the fact that they are the most distant
galaxies.

In the Type II objects, the $\co$ (3--2)/(1--0) ratio is practically
constant (at about 0.67) over the full range of infrared colors
$f_{60}/f_{100}$.  This mean $\co$ (3--2)/(1--0) ratio is, in fact,
identical to the mean of 0.66 found by Yao et et al. (2003) for
a sample of 60 infrared-luminous galaxies (with individual values
varying from from 0.22 to 1.72). Yao et al. also presented evidence
suggesting $X$ to be lower than the Solar Neighbourhood value by an
order of magnitude, making their sample more similar to our starburst
centers than to our compact galaxies.

{\it Type III (quiet compact)} consists of those galaxies where low
(2--1)/(1--0) line ratios $\sim 0.5$ suggest that on the whole little
is going on. Their molecular gas is not very dense, and may also not
be very warm. 2~Zw~40 is a member of this category, weak in $J=1-0$,
barely detected in $J=2-1$, and not in $J=3-2$. This class also apears
to include NGC~2730, NGC~3659, NGC~4532, NGC~6570, NGC~7732, and
IC~3521 from the sample observed by Albrecht et al. (2004). It would
be of interest to determine further line ratios in order to constrain
the physical conditions of the molecular gas, but the weakness of
emission in all but the $J$ = 1--0 transition makes this a difficult
task.

\section{Conclusions}

\begin{enumerate}

\item We have observed $\co$ line emission in the $J$=2--1 and
$J$=3--2 transitions from 11 compact (dwarf) galaxies. In
four cases, limited maps were made.

\item By combining our data with those from the literature, we
eliminated beam-dilution effects and established accurate line ratios
for the first three $\co$ rotational transitions in a sample of a
dozen objects, and limited the range of possible values in another
four. For six objects, $\co/\13co$ isotopical ratios were culled from
the literature.

\item Radiative-transfer (LVG) modelling shows that in most of the
observed galaxies warm molecular gas occurs with temperatures
typically $T_{k} = 50 - 150$ K as well as dense gas with $n(\h2) \geq
3000 \cc$. Models using two distinct molecular gas components produce
results different from and better than those obtained by modelling
only a single component.

\item Chemical modelling implies that in the observed galaxies only a
small fraction (typically $5\%$) of all gas-phase carbon is in the
form of CO, the remainder being in the form of neutral carbon ([CI])
and especially ionized carbon ([CII]). As a consequence, CO column
densities are quite low and of the order of $10^{16} \,\cm2$.

\item Molecular hydrogen column densities are high, of the order of
$10^{22} \,\cm2$, and confirm the large CO-to-H$_{2}$ conversion
factors, in the range $X = 10^{21} - 10^{22} \cm2/\kkms$, found
earlier for low-metallicity environments by different methods.

\item The CO spectroscopy of compact galaxies may be classified into
three different types. Type I is characterized by high rotational and
isotopic ratios, and reflects hot and dense molecular clouds dominated
by star-forming regions. Type II has lower ratios, in particular the
$\co$ (3--2)/(1--0) ratio is much lower than in Type I, and identical
to the mean found for infrared-luminous galaxies in general. Type III
has a low $\co$ (2--1)/(1--0) ratio indicative of not very dense and
possibly relatively cool molecular gas, and appears to represent
quiescent compact galaxies.

\end{enumerate}

\acknowledgements

We thank JACH personnel, in particular Fred Baas ($\dagger$), Remo
Tilanus and G\"oran Sandell for their help in obtaining the
observations discussed in this paper, and the referee, Ute Lisenfeld,
for critical remarks leading to improvements in the paper.

\end{document}